\documentclass[11pt,english]{article}
\date{}
\usepackage{appendix,braket}
\usepackage{authblk}
\usepackage{xcolor}
\usepackage[utf8x]{inputenc}
\usepackage{amsmath}
\usepackage{amsfonts}
\usepackage{amssymb}
\usepackage{graphicx,color}
\usepackage[english]{babel}
\usepackage{fontenc}
\usepackage{hyperref}
\def\be{\begin{equation}}
\def\ee{\end{equation}}

\begin{document}

\title{{\bf{Holographic subregion complexity of boosted black brane and Fisher information}}}
\author[1,3]{Sourav Karar\thanks{sourav.karar91@gmail.com}}
\author[2]{Rohit Mishra\thanks{mrohit996@gmail.com}}
\author[3]{Sunandan Gangopadhyay\thanks{sunandan.gangopadhyay@gmail.com}}
\affil[1]{Department of Physics, Government General Degree College, Muragachha 741154, Nadia, India}
\affil[2]{School of Physical Sciences, Indian Association for the Cultivation of Science,
2A $\&$ 2B Raja S.C. Mullick Road, Kolkata 700032, India}
\affil[3]{Department of Theoretical Sciences, S.N. Bose National Centre for Basic Sciences, Block-JD, Sector III, Salt Lake,  Kolkata 700106, India}

\maketitle
\begin{abstract}
\noindent In this paper, we have studied the holographic subregion complexity for boosted black brane for strip like subsystem. The holographic subregion complexity has been computed for a subsystem chosen along and perpendicular to the boost direction. We have observed that there is an asymmetry in the result due to the boost parameter which can be attributed to the asymmetry in the holographic entanglement entropy. The Fisher information metric and the fidelity susceptibility have also been computed using bulk dual prescriptions. It is observed that the two metrics computed holographically are not related for both the pure black brane as well as the boosted black brane. This is one of the main findings  in this paper and the holographic results have been compared with the results available in the quantum information literature where it is known that the two distances are related to each other in general.

\end{abstract}

\newpage

\section{Introduction}

The  discovery of gauge/gravity duality which relates conformal field theories living on the boundary of anti-de Sitter ($AdS$) spacetime to the bulk theory living in one extra spatial dimension has proved to be a remarkable progress in theoretical physics \cite{Witten:1998qj,Maldacena:1997re,Aharony:1999ti}. It has enabled us in getting deep insights in strongly coupled condensed matter systems and has also been the focus of recent developments in the field of string theory \cite{Calabrese:2004eu}-\cite{CasalderreySolana:2011us}.
Holographic computation of quantum information theoretic quantities such as entanglement entropy (EE) and complexity of a subsystem living on the boundary conformal field theory (CFT) has also been an intense area of research \cite{Ryu:2006bv}-\cite{Alishahiha:2018tep}. Investigations in these directions have led to
significant understanding of the basic laws governing such systems. It has been realized that small perturbations in the density matrix in the boundary field theory obey thermodynamic like relations which are similar to the black hole thermodynamical relations \cite{Bhattacharya:2012mi}-\cite{Mishra:2015cpa}. Interestingly such kind of relations have also been observed in the case of holographic complexity \cite{Karar:2017org,Karar:2018hvy}. The calculation prescription, for instance, for holographic EE involves the evaluation of the geometric area of  spatial extremal surfaces embedded in asymptotically $AdS$ spacetime whose boundary ends on the boundary of the subsystem at a fixed time \cite{Ryu:2006bv,Ryu:2006ef}.

In this paper, the computation of holographic subregion complexity (HSC)
for $(d+1)$-dimensional boosted black brane has been carried out for a strip like subsystem. The motivation of looking at such systems comes from the fact that there has been a lot of study where boosted brane solutions have provided a direct connection between the boundary theory and string theory. To be more precise, the AdS/CFT correspondence have been investigated in cases involving a $pp$-wave propagating along a particular direction in the world-volume of the classical $p$-brane configuration \cite{blau, mald}. Two distinct cases are known to arise here which depends on the configuration being BPS saturated or not. The effect of including the $pp$-wave in the non-BPS case turns out to be equivalent locally to a Lorentz boost given along the direction of propagation of the wave. It is to be noted that the validity of the equivalence is global if the direction of propagation of the $pp$-wave is uncompactified. However, if the direction along which the $pp$-wave propagates is wrapped on a circle, the equivalence is valid only locally. It is due to this reason that $p$-branes with $pp$-waves propagating on their world-volumes are refereed to as boosted $p$-branes. Further motivation of studying such systems comes from the fact that boosted AdS black brane backgrounds correspond to a thermal plasma uniformly boosted in a certain direction in the boundary theory \cite{Blanco:2013joa}.

Before proceeding with this computation, we would like to mention about the other proposals existing in the literature to compute the complexity holographically. The prescription to compute the holographic complexity (HC) was first proposed in \cite{sus1, sus2}. 
The proposal stated that the complexity of a state, measured in gates, is given by the volume of the Einstein-Rosen bridge (ERB) and reads
\begin{gather}
C_V (t_L , t_R)=\frac{V(t_L , t_R)}{R G}
  \label{ERB}
 \end{gather}
where $V$ is the spatial volume of the ERB. This volume is defined to be the maximum volume codimension one surface bounded by the CFT spatial slices at times $t_L$, $t_R$ on the two boundaries. The above formula is true up to a proportionality factor of order one which is unknown due to the problem of defining the gate complexity more precisely than this.

Yet another proposal for computing the HC of a
system exists in the literature. The content of the proposal is the following. The HC of a system can be obtained from the bulk action evaluated on the Wheeler-DeWitt patch \cite{Brown:2015lvg}
\begin{gather}
C_W=\frac{A(W)}{\pi \hbar}
  \label{new22}
 \end{gather}
where $A(W)$ is the action evaluated on the Wheeler-DeWitt
patch $W$ with a suitable boundary time. It is to be noted that in both the above proposals, the HC depends on the whole state of the physical system and therefore is not a property of a specific subsystem.

In this paper, we shall use the HSC proposal \cite{Mohsen}, which depends on the reduced state of the system.
It says that for subsystem $A$ in the boundary, if $V(\gamma)$ denotes the maximal codimension one volume enclosed by the codimension two minimal RT surface in the bulk, then the holographic subregion complexity can be calculated from the following formula 
\begin{gather}
C_V =\frac{V(\gamma)}{8\pi R G}
  \label{new1}
 \end{gather}
where $R$ is the radius of curvature of the spacetime.

With the thin strip approximation one may consider that the bulk extension only penetrates the ultra violet region of the spacetime under consideration. The boosted black brane geometry in the ultra violet limit can be considered as a perturbation around $AdS$ spacetime. In this paper, we have computed the HSC using the complexity-(RT) volume proposal in 
eq.(\ref{new1}) for the boosted black brane. In particular, we have carried out our computation upto both first and second orders in the perturbation parameter. Upto first order in perturbation, the HSC has been computed for the subsystem in both parallel and perpendicular directions of boost. We have then defined a ratio which can be identified as the holographic complexity asymmetry and compute this ratio. We have then extended our analysis to second order in perturbation. In this case we have computed the HSC for the subsystem which is perpendicular to the direction of boost. We have then moved on to compute holographically the Fisher information metric and the fidelity susceptibility for the boosted black brane. In this context, we would like to mention that there are two well known notions of distances in the quantum information literature, one is the Fisher information metric and the other is the Bures metric, also known as the fidelity susceptibility. It is also known in the quantum information literature that the two distances are the same for two infinitesimally close pure states \cite{PhysRevD.23.357} and other related cases \cite{PhysRevLett.72.3439,hubner1992explicit,sommers2003bures}. In general cases they are related \cite{Dominik}. We know that the proposed bulk duals of the boundary quantities should reproduce some of the properties of their boundary counterparts. For example holographic EE is shown to satisfy the subadditivity relations \cite{Headrick:2007km}-\cite{Wall:2012uf} and the entanglement first law \cite{Bhattacharya:2012mi}. Motivated by this result we would like to look in this paper its status in the holographic setup. The holographic computation of Fisher information metric is carried out using the proposal in \cite{Lashkari:2015hha} for both the pure black brane as well as the boosted black brane. We then follow the proposal in \cite{Banerjee:2017qti} to compute the Fisher information metric holographically. The proposal involves computing the change in subregion holographic complexity upto second order in perturbation about the pure $AdS$ spacetime and multiplying it by a dimensionless constant $C_d$. The constant is fixed by comparing it with the result obtained from the relative entropy \cite{Lashkari:2015hha,Blanco:2013joa,Beach:2016ocq}. The other notion of distance, the fidelity susceptibility has also been computed using the prescription given in \cite{MIyaji:2015mia}. It has been argued that the gravity dual of the fidelity susceptibility is approximately given by the volume of the maximal time slice in $AdS$ spacetime when the perturbation is exactly marginal. It was also generalized to incorporate mixed states also. The prescription was then applied to compute the fidelity susceptibility of the black brane. In this paper, this computation has been carried out for the boosted black brane.

Another aspect that has been looked at in this investigation is the following. Modifications to the holographic entanglement entropy (HEE) first law of thermodynamics have been obtained in $AdS$ spacetimes
carrying gauge charges \cite{Chakraborty:2014lfa,Mishra:2015cpa}. In particular, the boosted $AdS$ black branes have given rise to modifications of the first law of holographic entanglement thermodynamics (HET) \cite{Mishra:2015cpa,Mishra:2016yor}. It has been observed that the boosted black branes leads to an
asymmetry in the first law of HET. In this work, two cases were investigated, namely, strip subsystem parallel to the boost direction and also the other perpendicular to the boost direction. It was found that 
$\Delta S_{\bot} \geq \Delta S_{\parallel}$ and the entanglement asymmetry ratio was also computed. The asymmetry was found to be dependent on the boost parameter and was bounded from above. As mentioned in the earlier preceding paragraph that the computation of the HSC has also been done for two cases, namely, strip subsystem parallel to the boost direction and also the other perpendicular to the boost direction.
These studies also allow us to find asymmetry in the HSC for the two cases. Further, the investigations also puts some light on the dependence of the HSC with the holographic entanglement entropy. This dependence of the HSC on the HEE gives a possible reason for the asymmetry in the HSC since it was observed in \cite{Mishra:2016yor} that the HEE (for the strip subsystem parallel or perpendicular to the boost direction) has an asymmetry due to the difference in the entanglement pressure in the two directions.


This paper is organized as follows . In section (\ref{2}), computation of HSC for $(d+1)$- dimensional $AdS$ spacetime for a strip like subsystem 
has been done. The computation of HSC for $(d+1)$- dimensional boosted black brane has been carried out in section (\ref{3}). In section (\ref{4}), we given a detailed analysis of the Fisher information metric and the fidelity susceptibility for boosted black brane. We have concluded in section (\ref{5}). The paper also contains an appendix.


\section{Holographic subregion complexity for $(d+1)$- dimensional $AdS$ spacetime}{\label{2}}
In this section, we shall present a review of the computation of the HSC for
a strip like entangling surface in $(d+1)$- dimensional $AdS$ spacetime 
\cite{Ben-Ami:2016qex}.
The $AdS_{d+1}$ metric is given by
 \begin{gather}
  ds^2={-dt^2+{d{x_1}}^2+\cdots+{d{x_{d-1}}}^2+dz^2\over z^2}
 \end{gather}
where we have set the $AdS$ radius $R=1$. To compute the holographic subregion complexity 
we embed a strip like surface in this background given by $t=\text{constant},x_1=x_1(z)$.
The boundaries of the extremal bulk surface coincide with the two
ends of the interval $(-{l\over 2}\leq x_1\leq {l\over 2})$. 
The regulated size of the rest of the coordinates is taken to be large with $0\leq x_i\leq L_i$. 
The area of the strip like surface is given by

\begin{gather}
A_{(0)}=2V_{(d-2)}\int^{{z_*}^{(0)}}_{0} {dz\over z^{d-1}}\sqrt{1+{x_1'(z)}^2}
\end{gather}
 where $z_*^{(0)}$ is the turning point of the surface and $V_{(d-2)}=L_2 L_3 \cdots L_{d-1}$. 
 The minimal surface is obtained by minimizing the area functional. On minimizing we get
 \begin{gather}\label{minads}
  {x_1'(z)}={1\over\sqrt{\left({{z_*}^{(0)}\over z}\right)^{2(d-1)}-1}}~.
 \end{gather}
 The identification of the boundary $x_1(0)=l/2$ leads to the integral relation 
 \begin{gather}\label{lenads}
  {l\over 2}=\int_{0}^{{z_*}^{(0)}}{dz\over\sqrt{\left({{z_*}^{(0)}\over z}\right)^{2(d-1)}-1}}={z_*}^{(0)}\int_{0}^{1}t^{d-1}{dt\over\sqrt{1-t^{2(d-1)}}}={z_*}^{(0)} b_0
 \end{gather}
where $t={z\over {z_*}^{(0)}}$. The volume of the minimal surface is given by
\begin{gather}\label{volads}
 V_{(0)}=2 V_{(d-2)}\int_{\delta}^{{z_*}^{(0)}}{dz\over z^d}\int_{0}^{x_1(z)}dx_1(z)
\end{gather}
where $\delta$ is the UV cutoff. Now using eq. \eqref{minads} we can write eq. \eqref{volads} as
\begin{eqnarray}\label{V}
 V_{(0)}&=&2 V_{(d-2)}\int_{\delta}^{{z_*}^{(0)}}{dz\over z^d}\int_{z}^{{z_*}^{(0)}}({u\over {z_*}^{(0)}})^{d-1}{du\over\sqrt{1-({u\over {z_*}^{(0)}})^{2(d-1)}}} \notag \\
  &=&{ V_{(d-2)}\over(d-1) }{l\over\delta^{d-1}}-{ 2^{d-2}\pi^{(d-1)\over 2}\over(d-1)^2 }\left(\frac{ \Gamma \left(\frac{d}{2 d-2}\right)}{\Gamma \left(\frac{1}{2 (d-1)}\right)}\right)^{d-3}{ V_{(d-2)}\over l^{d-2}}
 \end{eqnarray}
where we have used eq. \eqref{lenads} in writing the second line of the above equation. 
Hence the HC for pure $AdS$ spacetime is given by
\begin{eqnarray}
 C_{(0)}&=&\frac{V_{(0)}}{8 \pi G_{(d+1)}}\notag \\
 &=&{ V_{(d-2)}\over 8 \pi G_{(d+1)}(d-1) }{l\over\delta^{d-1}}-{ 2^{d-2}\pi^{(d-1)\over 2}\over8 \pi G_{(d+1)}(d-1)^2 }\left(\frac{ \Gamma \left(\frac{d}{2 d-2}\right)}{\Gamma \left(\frac{1}{2 (d-1)}\right)}\right)^{d-3}{ V_{(d-2)}\over l^{d-2}}~.
\end{eqnarray}
Note that the first term in the holographic complexity is 
divergent (volume law) whereas the second term is finite. In the next section, we proceed to investigate the subregion holographic complexity of the boosted black brane.


\section{Holographic subregion complexity for  boosted $AdS_{d+1}$ black brane}{\label{3}}

The boosted $AdS_{d+1}$ black brane metric is given by
 \begin{equation}\label{bstmet}
ds^2={1\over z^2}\left( -{f dt^2\over K}+K (dy-\omega)^2
+dx_1^2+\cdots+dx_{d-2}^2+{dz^2 \over f}\right) 
\end{equation}
with
\begin{equation}
  K(z)=1+\beta^2\gamma^2{z^d\over z_0^d} ,~~~~~~~~~~f(z)=1-{z^d\over z_0^d}~.
\end{equation}
 where $0\le \beta\le 1$ is the boost parameter, $\gamma={1\over\sqrt{1-\beta^2}}$ and $z_0$ is the horizon of the 
 black brane. It is clear from the metric that the boost is taken along $y$ direction and the radius of curvature of
 $AdS$ spacetime has been set to one.
The Kaluza-Klein one form $\omega$ reads 
\begin{equation}
\omega={\beta^{-1}}(1-{1\over K}) dt.
\end{equation}
The metric \eqref{bstmet} has an anisotropy due to the boost along $y$ direction. This motivates us
to investigate the effect of anisotropy on subregion holographic complexity. To carry out this investigation, we compute the subregion 
complexity for two cases, firstly, for a strip along the direction of boost ($y$- direction) and secondly, for a strip
in the direction perpendicular to the boost ($x$- direction). 


\subsection{Strip parallel to the direction of boost}
To compute the HSC for a strip-like subregion we consider that
the boundaries of the extremal bulk surface  coincide with the two ends 
of the interval $-l/2 \le y \le l/2$ and $0\le x^i\le L_i$, with $L_i\gg l$.
The extremal surface is parametrized as $y =y(z)$.
Furthermore we have taken the strip
to be thin, so that the bulk extension can only penetrate the UV geometry.


Using the Ryu-Takayanagi prescription
the entanglement entropy for this strip like subsystem is given by
\begin{eqnarray}\label{entpara}
 S_\parallel  &\equiv& {Area(\gamma_A^{min})_{\parallel}\over 4G_{(d+1)}} \notag\\
&=&
 { V_{(d-2)}  \over 
2G_{(d+1)}}
\int^{z_\ast^{\parallel}}_{\delta}{dy\over z^{d-1}} \sqrt{K(z) + {({\partial_y z})^2\over{f(z)}} }
\end{eqnarray}  
where  
$G_{(d+1)}$ is $(d+1)$-dimensional Newton's constant, $\delta$ is the UV cutoff
and $V_{(d-2)} \equiv L_1 L_2 L_3 \cdots L_{d-2}$. As we are interested in computing the change in complexity
from pure $AdS$ spacetime, we can choose $L_i$ in such a way so that this $V_{(d-2)}$ has same
value as in pure $AdS$ case.
Here $z_\ast^{\parallel}$ is the turning point
of the extremal surface  inside the bulk geometry.

\noindent Now using the standard procedure of minimization
we obtain from eq. \eqref{entpara} the following expression for the extremal surface
\begin{eqnarray}
{dy\over dz}\equiv({z\over z_\ast^{\parallel}})^{d-1}{1 \over \sqrt{f(z) K(z)} 
 \sqrt{{K(z)\over K_\ast}- ({z\over z_\ast^{\parallel}})^{ 2d-2}}}  
\end{eqnarray} 
where  $K_\ast=K(z)|_{z=z_\ast^{\parallel}}$. To find an expression for the turning point in terms of the strip length
we make the identification $y(0)=l/2$. This gives 
\begin{equation}
\frac{l}{2}
= \int_0^{z_\ast^{\parallel}} dz \;
\left({z\over z_\ast^{\parallel}}\right)^{d-1}{1 \over \sqrt{f(z)\; K(z)} 
 \sqrt{{K(z)\over K_\ast}- ({z\over z_\ast^{\parallel}})^{ 2d-2}}}~. \notag\\
 \end{equation}
For small subsystem the turning point of the RT surface will be near to the $AdS$ boundary region $({z_\ast^{\parallel}}\ll{z_0})$. Now for finite boost we can evaluate the above integral by expanding it around the pure $AdS$ such that the condition
 \begin{equation}\label{lim}
  ({z_\ast^{\parallel}\over  z_0})^{d}\ll 1~~,~~\beta^2\gamma^2\left( \frac{z_\ast^{\parallel}}{z_0} \right)^d\ll 1
 \end{equation}
is always preserved. Under this approximation we can write the above integral as follows
\begin{eqnarray}\label{lpara} 
\frac{l}{2}&=& z_\ast^{\parallel} \int_0^{1} dt\;
t^{d-1}{1 \over  
 \sqrt{R}}[ 1+ {1\over 2}p^dt^d - {1\over 2}q^d t^d +{1\over 2}q^d {1-t^d\over R} +\cdots ]  \notag\\
&&\equiv
z_\ast^{\parallel} \left( b_0 +{1\over 2}(p^d b_1 - q^d b_1+q^d I_l)\right)  
+\cdots
\end{eqnarray}
where we have introduced  $R\equiv 1- t^{ 2d-2} $~,~$t=\frac{z}{z_\ast^{\parallel}}$, $q^d=\beta^2\gamma^2\left( \frac{z_\ast^{\parallel}}{z_0} \right)^d$,
$p=\frac{z_\ast^{\parallel}}{z_0}$
and the dots indicate terms of higher order in 
$({z_\ast^{\parallel}\over  z_0})^{d}$. The coefficients $b_0, b_1$, and $ I_l$ are provided in the appendix\eqref{betas}.

\noindent As we are using the metric \eqref{bstmet} to compute subregion HC keeping the strip length $l$ same
as in the case of pure $AdS$ spacetime, hence the turning point of the extremal surface will change. To express
the new turning point $z_\ast^{\parallel}$  in terms of ${z_*}^{(0)}$, which is the turning point in $AdS$ spacetime, we invert
eq. \eqref{lpara} and use eq. \eqref{lenads} to get
\begin{eqnarray}\label{plturn}
 z_{\ast}^{\parallel}&=&{l/2  \over  b_0 +{1\over 2}( p^d b_1-q^d b_1 + q^d  I_l)} 
 \simeq {{z_*}^{(0)} \over 1 + {1\over 2}( \bar{p}^d{b_1\over b_0}-\bar{q}^d {b_1 \over b_0} + \bar{q}^d  {I_l \over b_0})} 
\end{eqnarray}
where we have kept terms only up to $({z_\ast^{\parallel} \over  z_0})^{d}$ under the thin strip approximation and 
 $\bar{p}={{z_*}^{(0)}\over z_0}$ and $\bar{q}^d=\beta^2\gamma^2({{z_*}^{(0)}\over z_0})^d$.

\noindent Now the volume of the bulk extension under the RT minimal surface is given by
\begin{eqnarray}\label{volpara}
  V_{\parallel}&=&2 V_{(d-2)}\int^{z_\ast^{\parallel}}_{\delta}{dz\over z^d}\sqrt{{K(z)\over f(z)}}\int_z^{z_\ast^{\parallel}} dz
({u\over z_\ast^{\parallel}})^{d-1}{1 \over \sqrt{f(u) K(u)} 
 \sqrt{{K(u)\over K_\ast}- ({u\over z_\ast^{\parallel}})^{ 2d-2}}}\notag\\ 
&=&{2 V_{(d-2)}\over {z_\ast^{\parallel}}^{d-2}}\int^{1}_{\delta\over z_\ast^{\parallel}}{dt\over t^d}\sqrt{{K(t)\over f(t)}}\int_t^{1} dw \;
w^{d-1}{1 \over \sqrt{f(w) K(w)} 
 \sqrt{{K(w)\over K_\ast}- w^{ 2d-2}}}
 \end{eqnarray}
where $w={u\over z_\ast^{\parallel}}$, $K(w)=1+(wq)^d,~f(w)=1-(tp)^d$ . Now in the limit \eqref{lim} one
can make an expansion  of the functions ($K,f$) and keep terms up to linear order. This enables us to write the volume enclosed by the RT as a series around the pure $AdS$ volume.

Under these approximations we can expand the volume as
\begin{eqnarray}
V_{\parallel}&=&{2 V_{(d-2)}\over {z_*^{\parallel}}^{d-2}}\int^{1}_{\delta\over z_*^{\parallel}}{dt\over t^d}\left(1+{t^d\over 2}(p^d+q^d)\right)\int_t^{1} dw\;
w^{d-1}{1 \over \sqrt{f(w) K(w)} 
 \sqrt{{K(w)\over K_\ast}- w^{ 2d-2}}}\notag\\
&=&{2 V_{(d-2)}\over {z_*^{\parallel}}^{d-2}}\int^{1}_{\delta\over z_*^{\parallel}}{dt\over t^d}\int_t^{1} dw\;
w^{d-1}{1 \over \sqrt{f(w)K(w)} 
 \sqrt{{K(w)\over K_\ast}- w^{ 2d-2}}}\notag\\&&+{ V_{(d-2)}\over {z_*^{\parallel}}^{d-2}}\left(p^d+q^d\right)\int^{1}_{\delta\over z_*^{\parallel}}dt\int_t^{1} dw\;
w^{d-1}{1 \over \sqrt{f(w)K(w)} 
 \sqrt{{K(w)\over K_\ast}- w^{ 2d-2}}}~.\notag\\
 \end{eqnarray}
After evaluating these straight forward integrals, we use eq. \eqref{plturn} to obtain the minimal volume $V_{\parallel}$
in terms of the minimal volume of pure $AdS$ spacetime (eq.\eqref{V}) to get
\begin{eqnarray}
  V_{\parallel}&=&V_{(0)}- { V_{(d-2)} \bar{p}^d \over (d-1){{z_*}^{(0)}}^{d-2}} \left({(d-2)\pi b_1 \over 2(d-1) b_0^2} + (2-d)c_0 \right) \notag \\
&&-{ V_{(d-2)} \bar{q}^d \over (d-1){{z_*}^{(0)}}^{d-2}} \left({(d-2)\pi  \over 2 (d-1)^2 b_0}(\frac{2b_1}{b_0}-1) +c_2 -c_0 d \right)
\end{eqnarray}
where we have kept terms upto linear order in $\bar{p}^d$ and $\bar{q}^d$. We can recast 
the change in volume using eq. \eqref{lenads} in terms of the length $l$ of the strip as
\begin{eqnarray}
\Delta V_\parallel &=&V_\parallel-V_{(0)} \notag\\
&=&-{ V_{(d-2)}{l}^2\over 4 {b_0}^2 (d-1) {z_0}^d}\left[\left({(d-2)\pi b_1\over 2 (d-1){b_0}^2}+(2-d)c_0\right) \right.\notag \\
&& \left. +\beta^2\gamma^2\left( {(d-2)\pi  \over 2 (d-1)^2 b_0}(\frac{2b_1}{b_0}-1) +c_2 -c_0 d\right)\right] ~.
\end{eqnarray}

\noindent Hence the change in complexity for a strip parallel to the direction of boost is given by
\begin{eqnarray}
 \Delta C^{(1)}_\parallel &\equiv& \frac{\Delta V_\parallel }{8 \pi G_{(d+1)}} \notag \\
 &=& -{ V_{(d-2)}{l}^2\over 32 \pi  G_{(d+1)}{b_0}^2 (d-1) {z_0}^d} \notag \\
 && \times \left[\left({(d-2)\pi b_1\over 2 (d-1){b_0}^2}+(2-d)c_0\right)+\beta^2\gamma^2\left( {(d-2)\pi  \over 2 (d-1)^2 b_0}(\frac{2b_1}{b_0}-1) +c_2 -c_0 d\right)\right]. \notag\\
\end{eqnarray}
It can be clearly seen that the change in holographic complexity depends on the boost parameter. Note that in the $\beta\rightarrow 0$ limit, the result agrees with the pure black brane result obtained in \cite{Ben-Ami:2016qex}.

This can be recast in the following form
\begin{equation}{\label{c1para}}
  \Delta C^{(1)}_\parallel = -\frac{\pi (d-2)}{2(d-1)^3 b_0^2}\left[\Delta S_\parallel -\frac{b_0^2}{(d+1)b_1^2 }\Delta S_\bot\right]
\end{equation}
where the expression for $\Delta S_\parallel$ and $\Delta S_\bot$ are given by \cite{Mishra:2016yor},
\begin{eqnarray}
\Delta S_\parallel &=&\frac{V_{(d-2)} l^2 b_1 (d+1)}{32G_{(d+1)} b_0^2 z_0^d}\left(\frac{d-1}{d+1}+\frac{2}{d+1}\beta^2\gamma^2 \right)\notag \\
\Delta S_\bot&=& \frac{V_{(d-2)} l^2 b_1 (d+1)}{32G_{(d+1)} b_0^2 z_0^d} \left(\frac{d-1}{d+1} +\beta^2\gamma^2  \right)
\end{eqnarray}
are the change in entanglement entropies up to first order in perturbation.

\noindent The interesting point to note in the above result is that the change in holographic complexity in the parallel direction contains information of the changes in the holographic entanglement entropy in both the parallel and the perpendicular directions of the boost with the respect to the subsystem.

\subsection{Strip perpendicular to the direction of boost}
In this subsection, we essentially follow the analysis similar to the earlier section to compute the HSC of the strip like subsystem with the strip being perpendicular to the direction of boost. Using the Ryu-Takayanagi prescription the entanglement entropy for this strip like subsystem is as follows
\begin{eqnarray}\label{entperp}
 S_\bot  &\equiv& {Area(\gamma_A^{min})_{\bot}\over 4G_{(d+1)}} \notag\\
&=&
 { V_{(d-2)}  \over 
2G_{(d+1)}}
\int^{z_\ast^{\bot}}_{\delta}{dz\over z^{d-1}} \sqrt{K(z)} 
\sqrt{{1\over {f(z)}}  +({\partial_z x^1})^2}~.
\end{eqnarray}  
Here $z_\ast^{\bot}$ is the turning point
of the extremal surface  inside the bulk geometry. Now using the standard procedure of minimization
we obtain from eq. \eqref{entperp} the following expression for extremal surface
\begin{eqnarray}
{dx^1\over dz}=\left({z\over z_\ast^{\bot}}\right)^{d-1}{1 \over \sqrt{f(z)} 
 \sqrt{{K(z)\over K_\ast}- ({z\over z_\ast^{\bot}})^{2d-2}}} 
\end{eqnarray} 
where  $K_\ast=K(z)|_{z=z_\ast^{\bot}}$. To find an expression for the turning point in terms of the strip length
we make the identification $x^1(0)=l/2$. This yields  
\begin{equation}
{l \over 2}=\int_0^{z_\ast^{\bot}} dz
\left({z\over z_\ast^{\bot}}\right)^{d-1}{1 \over \sqrt{f(z)} 
 \sqrt{{K(z)\over K_\ast}- ({z\over z_\ast^{\bot}})^{ 2d-2}}} \notag\\
\end{equation}
where we have taken the same subsystem size as in the parallel case and is assumed to be small. Hence the turning point will lie near the asymptotic region $(z_\ast^{\bot}\ll z_0)$. Thus for finite boost we can expand the above integral around pure AdS preserving the following condition
\begin{equation}\label{limperp}
  ({z_\ast^{\bot}\over  z_0})^{d}\ll 1~~,~~\beta^2\gamma^2\left( \frac{z_\ast^{\bot}}{z_0} \right)^d\ll 1
 \end{equation}
 Thus in this limit, expanding the above integral gives
 \begin{eqnarray}\label{lperp} 
 {l \over 2}&=&z_\ast^{\bot} \int_0^{1} dt\;
t^{d-1}{1 \over  
 \sqrt{R}}[ 1+ {1\over 2}x^dt^d +{1\over 2}y^d {1-t^d\over R} +\cdots ]  \notag\\
 && = z_\ast^{\bot}\left( b_0 +{1\over 2}(x^d b_1 +y^d I_l)\right)  
+\cdots
\end{eqnarray}
where we have introduced  $R\equiv 1- t^{ 2d-2} $, $t={z\over z_\ast^{\bot}}$, $y^d=\beta^2\gamma^2({z_*^{\bot}\over z_0})^d$, $x=({z_*^{\bot}\over z_0})$
and the dots indicate terms of higher order in 
$({z_\ast^{\bot}\over  z_0})^{d}$. 

\noindent To express
the new turning point $z_\ast^{\bot}$  in terms of ${z_*}^{(0)}$,  we invert
eq. \eqref{lperp} and use eq. \eqref{lenads} to get
\begin{eqnarray}\label{pturn}
 z_{\ast}^{\bot}&=&{l/2  \over  b_0 +{1\over 2}( x^d b_1 + y^d  I_l)} 
 \simeq {{z_*}^{(0)} \over 1 + {1\over 2}( \bar{x}^d{b_1\over b_0} + {\bar{y}^d \over  b_0}I_l)} 
\end{eqnarray}
where we have kept the terms only upto $({z_\ast^{\bot}\over  z_0})^{d}$ under the thin strip approximation. Note that the two turning points, the perpendicular and the parallel, reduces to the same result in the $\beta\rightarrow 0$ limit.

\noindent Now the volume of the bulk extension under RT minimal surface is given by
\begin{eqnarray}\label{volperp}
 V_{\bot}&=&2 V_{(d-2)}\int^{z_*^{\bot}}_{\delta}{dz\over z^d}\sqrt{{K(z)\over f(z)}}\int_z^{z_\ast^{\bot}} dz
({u\over z_\ast^{\bot}})^{d-1}{1 \over \sqrt{f(u)} 
 \sqrt{{K(u)\over K_\ast}- ({u\over z_\ast^{\bot}})^{ 2d-2}}}\notag\\ 
&=&{2 V_{(d-2)}\over {z_*^{\bot}}^{d-2}}\int^{1}_{\delta\over z_*^{\bot}}{dt\over t^d}\sqrt{{K(t)\over f(t)}}\int_t^{1} dw \;
w^{d-1}{1 \over \sqrt{f(w)} 
 \sqrt{{K(w)\over K_\ast}- w^{ 2d-2}}}
 \end{eqnarray}
Once again in the limit ${z_*^{\bot}\over z_0}\ll 1$, one can make an asymptotic expansion  of the defining functions ($K,f$) in terms of this parameter up to linear order. 
Under these approximations we can write the volume as
\begin{eqnarray}
V_{\bot}&=&{2 V_{(d-2)}\over {z_*^{\bot}}^{d-2}}\int^{1}_{\delta\over z_*^{\bot}}{dt\over t^d}\left(1+{t^d\over 2}(x^d+y^d)\right)\int_t^{1} dw \;
w^{d-1}{1 \over \sqrt{f(w)} 
 \sqrt{{K(w)\over K_\ast}- w^{ 2d-2}}}\notag\\
&&={2 V_{(d-2)}\over {z_*^{\bot}}^{d-2}}\int^{1}_{\delta\over z_*^{\bot}}{dt\over t^d}\int_t^{1} dw\;
w^{d-1}{1 \over \sqrt{f(w)}
 \sqrt{{K(w)\over K_\ast}- w^{ 2d-2}}}\notag\\&&+{ V_{(d-2)}\over {z_*^{\bot}}^{d-2}}\left(x^d+y^d\right)\int^{1}_{\delta\over z_*^{\bot}}dt\int_t^{1} dw
w^{d-1}{1 \over \sqrt{f(w)} 
 \sqrt{{K(w)\over K_\ast}- w^{ 2d-2}}}.
 \end{eqnarray}
Evaluating these straight forward integrals we use eq. \eqref{pturn} to obtain the minimal volume
in terms of the minimal volume of pure $AdS$ spacetime. This leads to
\begin{gather}
  V_{\bot}=V_{(0)}-{ V_{(d-2)}\bar{x}^d\over (d-1) {{z_*}^{(0)}}^{d-2}}\left({(d-2)\pi b_1\over 2 (d-1){b_0}^2}+c_0\right)
  -{ V_{(d-2)}\bar{y}^d\over(d-1){{z_*}^{(0)}}^{d-2}}\left({(d-2)\pi I_l\over2 (d-1){b_0}^2}+c_2\right)\notag \\
  +{ V_{(d-2)}c_0\over {{z_*}^{(0)}}^{d-2}}\left(\bar{x}^d+\bar{y}^d\right)
\end{gather}
where we have kept terms up to linear order in $\bar{x}^d$ and $\bar{y}^d$. In terms of the length $l$ of the 
strip, the change in volume can be recast in the form
\begin{eqnarray}
\Delta V_\bot &\equiv&V_\bot-V_{(0)} \notag\\
&=&-{ V_{(d-2)}{l}^2\over 4 {b_0}^2 (d-1) {z_0}^d}
 \left[\left({(d-2)\pi b_1\over 2 (d-1){b_0}^2}+(2-d)c_0\right)\right. \notag \\
 && \left. +\beta^2\gamma^2\left({(d-2)\pi I_l\over2 (d-1){b_0}^2}+
 c_2-(d-1)c_0\right)\right]~.
\end{eqnarray}
Hence the change in complexity for a strip perpendicular to the direction of boost is given by
\begin{eqnarray}
 \Delta C^{(1)}_\bot &=& \frac{\Delta V_\bot}{8 \pi G_{(d+1)}} \notag \\
 &=& -{ V_{(d-2)}{l}^2\over 32 \pi  G_{(d+1)}{b_0}^2 (d-1) {z_0}^d} \notag \\
 && \times \left[\left({(d-2)\pi b_1\over 2 (d-1){b_0}^2}+(2-d)c_0\right)+\beta^2\gamma^2\left({(d-2)\pi I_l\over2 (d-1){b_0}^2}+
 c_2-(d-1)c_0\right)\right] \notag\\
\end{eqnarray}
which can be recast in the following form
\begin{equation}{\label{c1perp}}
 \Delta C^{(1)}_\bot = \Delta C^{(1)}_\parallel -\frac{ V_{(d-2)} l^2 \beta^2\gamma^2 c_0}{32\pi G_{(d+1)} (d-1)b_0^2 z_0^d} \left[ 1+(d-2)(d+1)\frac{b_1^2}{b_0^2}\right]~.
\end{equation}
Note that there is an asymmetry in the holographic subregion complexities in both the directions which owes its origin to the boost parameter. In the $\beta \rightarrow 0$ limit, this asymmetry vanishes and reassuringly both the changes agree with each other. 

\noindent The above expression for $\Delta C^{(1)}_\bot$ can also be recast in a form similar to eq.(\ref{c1para}). This reads
\begin{equation}{\label{c1perpend}}
 \Delta C^{(1)}_\bot =\frac{1}{2(d-1)^3}\left[\frac{\Delta S_{\parallel}}{(d+1)b_{1}^2}-\Delta S_{\bot}\left(\frac{d-2}{b_{0}^2}-\frac{d-3}{(d+1)b_{1}^2}\right)\right] ~.
 \end{equation}

\noindent We now proceed to investigate the asymmetry in the result in HSC for the parallel and perpendicular directions of the boost with respect to the subsystem size. For this 
let us define a quantity 
\begin{equation}
\mathcal{R}_C= \frac{\Delta C^{(1)}_\bot -\Delta C^{(1)}_\parallel}{\Delta C^{(1)}_\bot +\Delta C^{(1)}_\parallel}~.
\end{equation}
This can be called the holographic subregion complexity ratio. To understand the effect of anisotropy on the HSC, we use eq.(s) (\ref{c1para}, \ref{c1perpend}) to get
\begin{equation}
\mathcal{R}_C=\frac{\left[\frac{2-d}{2(d-1)^3 b_{0}^2}-\frac{1}{2(d-1)^3 (d+1)b_{1}^2}\right]\mathcal{A}}{\frac{2-d}{2(d-1)^3 b_{0}^2}+\frac{\mathcal{R}+2d-5}{2(d-1)^3 (d+1)b_{1}^2}}
\label{new12}
\end{equation}
where $\mathcal{R}$ and $\mathcal{A}$ are given by \cite{Mishra:2016yor}
\begin{equation}
\mathcal{R}=\frac{\Delta S_{\parallel}}{\Delta S_{\bot}}=\frac{1+\frac{2}{d-1}\beta^2 \gamma^2}{1+\frac{d+1}{d-1}\beta^2 \gamma^2}
\label{new2}
\end{equation}
\begin{eqnarray}
\mathcal{A}&=&\frac{\Delta S_{\bot}-\Delta S_{\parallel}}{\Delta S_{\bot}+\Delta S_{\parallel}}=\frac{1-\mathcal{R}}{1+\mathcal{R}}\nonumber\\
&=&\frac{\beta^2 \gamma^2}{2 +\frac{d+3}{d-1}\beta^2 \gamma^2}~.
\label{new3}
\end{eqnarray}
It is easy to check that $\mathcal{R}_C \ge 0$ for $\beta \ge 0$. Hence, there is an asymmetry in the holographic subregion complexities for the perpendicular and parallel directions of the boost with respect to the system. This asymmetry in the holographic subregion complexities owes its origin to the asymmetry in the holographic entanglement entropies in the perpendicular and the parallel directions. The key to this holographic entanglement entropy asymmetry can be related to the unequal entanglement pressure \cite{Mishra:2016yor}.
Studies of holographic entanglement entropy in spatially anisotropic field theory also showed such asymmetry owing its origin to the fact that the pressure in one direction is different from the others \cite{rahimi}. This is also consistent with the results obtained using the complexity equals action proposal \cite{HosseiniMansoori:2018gdu}.

It is important to note from eq.\eqref{new12} that the minimum value of $R_C$ is zero which corresponds to $\beta=0$. Further we note from eq.\eqref{new12} that the maximum value of $R_C$ will depend on the maximum value of $\mathcal{A}$ and $\mathcal{R}$. Their maximum value is obtained by taking the simultaneous limit $\beta\to 1$ and $z_0\to\infty$ keeping the ratio ${\beta^2\gamma^2\over z^d_0}={1\over z^d_I}=\text{fixed}$ such that the perturbative expression remains valid. In this simultaneous limit the boosted black brane geometry reduces to the AdS $pp$-wave background 
\begin{eqnarray}\label{bstpar}
&&ds^2={L^2\over z^2}\left( -K^{-1}{ dt^2}+K (dy-(1-K^{-1}) dt)^2
+dx_1^2+\cdots+dx_{d-2}^2+{dz^2  }\right)\notag\\ 
\end{eqnarray}
where $K(z)=1+{z^d\over z^d_I}$. For this geometry the entanglement pressure along the wave direction only is non-zero while the entanglement pressure in all other directions vanishes. Thus the pressure (difference) asymmetry is maximum. In this background the maximum values of $\mathcal{R}={2\over d+1}$ and $\mathcal{A}={d-1\over d+3}$ is acheived. Thus the difference in entanglement pressure in the CFT is the source of the asymmetry in complexity.

In the $\beta \rightarrow 0$ limit $\mathcal{R}_C =0$. The $\beta \rightarrow 0$ limit implies that the complexity is same ($\Delta C^{(1)}_\bot = \Delta C^{(1)}_\parallel$) in both the parallel or perpendicular directions of the subsystem.




\subsection{Holographic subregion complexity upto second order in perturbation}

In this section we have computed the HSC for strip like subregion in boosted black brane \eqref{bstmet}
background with the perturbation up to second order in $({z_*\over z_0})^d$ and $\beta^2\gamma^2({z_*\over z_0})^d$ 
around the pure $AdS$ background. 
The strip has been chosen to be in a direction perpendicular to the direction of boost. 
As we are interested in second order perturbation, therefore the expression for the turning point 
and volume will receive some corrections. As the volume under minimal surface depends
upon the turning point of the minimal surface, hence we shall first compute the change in turning point. To begin 
with let us first compute the length $l$ of the subsystem perturbatively in the same limit as in eq.no \eqref{limperp} up to second order.
The expression for length of the subsystem  when the strip is perpendicular to the direction of
boost is given by
\begin{eqnarray}\label{lperp2}
{l\over 2} &=&\int_0^{z_*} {dz \over \sqrt{1-({z\over z_0})^d}}\frac{(z/ z_*)^{d-1}}{\sqrt{{K(z)\over K_*}-(z/ z_*)^{2(d-1)}}}\notag\\
&=& z_*\left[ \int_0^1 dt{t^{d-1} \over \sqrt{R}} +\frac{x^d}{2} \int_0^1 dt{t^{d-1} \over \sqrt{R}}\left(t^d +\beta^2 \gamma^2 \frac{1-t^d}{R} \right) \right. \notag \\
&&\left. + x^{2d} \int_0^1 dt{t^{d-1} \over \sqrt{R}} \left(\frac{3}{8}t^{2d} + \frac{\beta^2 \gamma^2}{4}\frac{t^d(1-t^d)}{R}+\beta^4 \gamma^4 \left( \frac{3}{8}\frac{(1-t^d)^2}{R^2}-\frac{1}{2}\frac{1-t^d}{R}\right) \right) \right] \notag \\
&=& z_*\left[b_0 + \frac{x^d}{2} (b_1 + \beta^2 \gamma^2 I_l)+ x^{2d}({3 \over 8}b_2 +J_l)\right],
\end{eqnarray}
where the coefficients $b_0$, $b_1$, $I_l$ and $J_l$ are given in the appendix\eqref{betas}. In order to
express this turning point $z_*$ in terms of the turning point $z_*^{(0)}$of pure $AdS$ background,
we use eq. \eqref{lenads} to get
\begin{equation}
 z_*={z_*^{(0)} \over 1 + \frac{1}{2b_0}(b_1+ \beta^2 \gamma^2 I_l)\bar{x}^d + (\frac{3b_2}{8b_0}+\frac{J_l}{b_0} -d(\frac{b_1+ \beta^2 \gamma^2 I_l}{2b_0})^2)\bar{x}^{2d}}
\end{equation}
where $\bar{x}= z_*^{(0)}/z_0$. Now the expression for volume is given by
\begin{eqnarray}
 V\simeq \frac{2 V_{(d-2)}}{z_*^{d-2}} \int_{\frac{\delta}{z_*}}^1 \frac{dt}{t^d}\left(1+\frac{x^d+y^d}{2} t^d +(\frac{3}{8}x^{2d} +\frac{x^d y^d}{4}-\frac{y^{2d}}{8})t^{2d} \right)\notag \\
 \times \int_t^{1} dw \frac{1}{\sqrt{f(w)}}
 \frac{w^{d-1}}{\sqrt{\frac{K(w)}{K_*}-w^{2(d-1)}}}~.
\end{eqnarray}
After some lengthy calculations, we obtain the volume from the above expression to be
\begin{gather}
 V=V_{(0)} -\frac{V_{(d-2)}\bar{x}^d}{(d-1)\bar{z}_*^{d-2}} \left(\frac{d-2}{d-1}\frac{\pi b_1}{2b_0^2}+(2-d)c_0\right) -\frac{V_{(d-2)}\bar{y}^d}{(d-1)\bar{z}_*^{d-2}}\left(\frac{d-2}{d-1}\frac{\pi I_l}{2b_0^2}+c_2-(d-1)c_0\right) \notag\\
 -\frac{V_{(d-2)}\bar{x}^{2d}}{\bar{z}_*^{d-2}}v_{00}
 -\frac{V_{(d-2)}\bar{x}^{d}\bar{y}^d}{\bar{z}_*^{d-2}}v_{01}
 +\frac{V_{(d-2)}\bar{y}^{2d}}{\bar{z}_*^{d-2}} v_{11}
\end{gather}
where $V_{(0)}$ is the volume under RT surface for pure $AdS$ given in eq. \eqref{V} with
\begin{eqnarray}
 v_{00}&=& \left(\frac{3\pi b_2}{8b_0^2}\frac{d-2}{(d-1)^2}-\frac{\pi b_1^2}{8b_0^3}\frac{(d-2)(d+3)}{(d-1)^2}+\frac{c_0b_1}{b_0}\frac{d-2}{d-1}-\frac{c_1}{2}\frac{d^2-4}{d^2-1}\right) \notag \\
v_{01}&=& \left(\frac{b_1}{b_0}(c_0-\frac{c_2}{d-1})-(\frac{c_3}{2}+\frac{(d+2)c_1}{2(d+1)})+\frac{d-2}{d-1}\frac{c_0 I_l}{b_0}+\frac{2K_1}{d-1} \right.\notag \\
&&\left. +\frac{d-2}{(d-1)^2}\frac{\pi J_1}{b_0^2} -\frac{(d-2)(d+3)}{(d-1)^2}\frac{\pi b_1 I_l}{4b_0^3}\right) \notag \\
v_{11}&=&  \left(\frac{c_3}{2}-\frac{c_1}{4(d+1)}+\frac{2K_2}{d-1}-(c_0 - \frac{c_2}{d-1})\frac{I_l}{b_0}-\frac{d-2}{(d-1)^2}\frac{\pi J_2}{b_0^2}+\frac{(d-2)(d+3)}{(d-1)^2}\frac{\pi  I_l^2}{8b_0^3}\right)~.\notag\\
 \end{eqnarray}
Now using the expression given in the earlier section for subregion HC for strip perpendicular to the direction 
of boost with perturbation
up to first order, we write the HSC upto second order in the perturbation to be
\begin{eqnarray}{\label{c2perp}}
 \Delta C=\Delta C^{(1)}_\bot +\Delta C^{(2)}_\bot
\end{eqnarray}
where
\begin{eqnarray}{\label{c2perp1}}
\Delta C^{(2)}_\bot =-\frac{V_{(d-2)} l^{d+2}}{8\pi G_{(d+1)}z_0^{2d} (2b_0)^{d+2}}\left[v_{00} + \beta^2\gamma^2 v_{01} -  \beta^4\gamma^4 v_{11}\right]~.
\end{eqnarray}
This is the second order change in the holographic complexity.
We will use this expression in the next section to calculate the Fisher Information metric. It is important to note that the boosted black brane is a stationary spacetime. For such spacetimes one should use the covariant HRT proposal instead of the static RT proposal. However it can be shown that at first order of the perturbative expansion it is sufficient to take the $t=constant$ slicing. At first order, the sole contribution comes from the metric perturbations \cite{Nozaki:2013vta}-\cite{Ghosh:2016fop}. Deviations of the minimal surface only contribute at the second order. Thus at second order one cannot work with the same $t=constant$ embedding for stationary asymptotically AdS spactimes. But as shown in \cite{Ghosh:2017ygi} one can still work with the $t=constant$ slice in the boosted black brane spacetime but only for the perpendicular case. This is due to the fact that the minimal surface still remains in the same time slice. The deviations contribute in other spatial directions.

\section{Fisher information metric and Fidelity susceptibility}{\label{4}}
In this section, we shall compute the Fisher information metric and the fidelity susceptibility
for the boosted black brane from the proposals existing in the literature.
Before we begin our analysis we would like to briefly mention about the quantities in
the context of quantum information theory. Two well known notion of distance between two quantum states exist in the literature. One is the Fisher information metric and the other is the Bures metric or fidelity susceptibility. The Fisher information metric is defined as \cite{Banerjee:2017qti}
\begin{equation}
G_{F,\lambda \lambda}=\langle \delta \rho\; \delta \rho\rangle_{\lambda \lambda}^{(\sigma)}=
\frac{1}{2}tr \left(\delta \rho \frac{d}{d(\delta \lambda)} \log(\sigma + \delta \lambda \delta \rho)|_{\delta \lambda=0} \right)
\label{FI}
\end{equation}
where $\delta \rho$ is a small deviation from the density matrix $\sigma$.

\noindent A second notion of distance between two states is known as fidelity susceptibility and reads
\begin{equation}\label{fidelity}
G_{\lambda \lambda }= \partial^2_{\lambda} F ;~~~~F= tr\sqrt{\sqrt{\sigma_{\lambda}} \rho_{\lambda + \delta \lambda} \sqrt{\sigma_{\lambda}}}
\end{equation} 
where $\sigma $ and $\rho$ are the initial and final density matrices, $F$ is called the fidelity.

\noindent The first holographic computation of the Fisher information metric was carried out
in \cite{Lashkari:2015hha}, with the Fisher information  metric defined as 
\begin{equation}{\label{fisher}}
G_{F,mm}=\frac{\partial^2 }{\partial m^2}S_{rel}(\rho_m \parallel \rho_0);~~~~~~~ S_{rel}(\rho_m \parallel \rho_0)=\Delta \langle H_{\rho_0} \rangle-\Delta S
\end{equation}
where $m$ is a perturbation parameter, $\Delta \langle H_{\rho_0} \rangle$ is the change in modular Hamiltonian
and $\Delta S$ is the change in entanglement entropy from the vacuum state. It has been shown in \cite{Lashkari:2015hha}
that at first order in $m$ the relative entropy vanishes(Entanglement First Law) and in second order in $m$ the
relative entropy is given by $S_{rel}=-\Delta S^{(2)}$. With this basic background in place we first compute the Fisher information metric for the black brane.

\noindent The inverse of the lapse
function can be written as
\begin{equation}
\frac{1}{f(z)}=\frac{1}{1-{z^d\over z_0^d}}=1+mz^d +m^2z^{2d}+\cdots
\end{equation}
where $m=1/z_0^d$, which is the perturbation parameter in the bulk. In terms of this parameter
the change in area of the minimal surface upto second order is given by \cite{Blanco:2013joa}
\begin{equation}
A-A_{0}= \left[\frac{V_{(d-2)}a_1 l^2}{4 b_0^2}\frac{d-1}{d+1}m +
\frac{V_{(d-2)} a_1 h_0 l^{d+2}}{(2b_0)^{d+2}} m^{2}\right]
\end{equation}
where
\begin{equation}
h_0 = \frac{d-1}{d+1}\left(-\frac{b_1}{2b_0}+\frac{3(d+1)}{4(2d+1)}\frac{a_2}{a_1}  \right)~.
\end{equation}
The relative entropy in this case becomes
\begin{equation}
S_{rel}= -\frac{1}{4G_{(d+1)}} \left[
\frac{V_{(d-2)} a_1 h_0 l^{d+2}}{(2b_0)^{d+2}} m^{2}\right]~.
\end{equation}
It is to be noted that $h_0$ has negative values for all $d$. Hence 
$S_{rel}$ is a positive quantity.

\noindent From eq. \eqref{fisher}, the Fisher information metric therefore reads
\begin{equation}\label{fisher0}
G_{F,mm}=\frac{\partial^2 }{\partial m^2} S_{rel} = -\frac{V_{(d-2)} a_1 h_0 l^{d+2}}{2 G_{(d+1)}(2b_0)^{d+2}} ~.
\end{equation} 

\noindent In \cite{Banerjee:2017qti}, a proposal for computing the above quantity was given.
The proposal is to consider the difference  of two volumes yielding a finite expression
\begin{equation}
 \mathcal{F}=C_d(V_{(m^2)}-V_{(0)})
\end{equation}
where $ V_{(m^2)}$ is evaluated for a second order fluctuation about $AdS$ spacetime. $C_d$ is a dimensionless constant which 
cannot be fixed from the first principles of the gravity side. We shall now apply this proposal to compute
the Fisher information metric for the black brane.
The change in volume under Ryu-Takayanagi minimal surface at second order in perturbation takes the form (for $\beta =0$ case)
\begin{equation}
V-V_{(0)}= -\left[{ V_{(d-2)}{l}^2\over 4{b_0}^2 (d-1) }m v_0
+\frac{V_{(d-2)} l^{d+2}}{ (2b_0)^{d+2}} m^2 v_{00}\right]~.
\end{equation}
The holographic dual of Fisher information metric is now defined as
\begin{equation}{\label{deb}}
G_{F,mm}=\partial_m^2 \mathcal{F};~~~~~~\mathcal{F}= C_d  \left(V-V_{(0)} \right)
\end{equation}
with the constant $C_d$ to be determined by requiring that the holographic dual Fisher information
metric from the above equation must agree with that obtained from the relative
entropy \eqref{fisher0}. The constant $C_d$ is therefore given by
\begin{equation}{\label{cd0}}
C_d=\frac{h_0 a_1}{4G_{(d+1)} v_{00}} ~.
\end{equation} 

\noindent On the other hand the relative entropy for boosted black brane ($\beta \neq 0$) is given by \cite{Mishra:2015cpa}
\begin{equation}
 S_{rel} =-\frac{1}{4G_{(d+1)}}\left[ \left(\frac{l}{2b_0}\right)^{d+2} \left(h_0 +h_1 \beta^2 \gamma^2 +h_2 \beta^4 \gamma^4 \right)m^2\right]
\end{equation}
where 
\begin{eqnarray}
 h_1 &=& \left(-\frac{b_1}{b_0}+ \frac{a_2}{2 a_1} \right) \notag \\
 h_2 &=& \frac{d+1}{d-1}\left( -\frac{b_1}{2b_0} +\frac{3a_2}{4a_1(d+1)}\right) ~.
\end{eqnarray}
It is to be noted that $h_1$ and $h_2$ have negative values for all $d$. Hence 
$S_{rel}$ is a positive quantity. Thus the Fisher information metric reads
\begin{equation}{\label{fisher1}}
 G_{F,mm}=\frac{\partial^2 }{\partial m^2} S_{rel} = -\frac{1}{2G_{(d+1)}}\left(\frac{l}{2b_0}\right)^{d+2}
 \left(h_0 +h_1 \beta^2 \gamma^2 +h_2 \beta^4 \gamma^4 \right)~.
\end{equation}

\noindent Now the change in volume under Ryu-Takayanagi minimal surface at second order in perturbation takes the form (for $\beta \neq 0$ case)
\begin{gather}
 V-V_{(0)}= -\frac{V_{(d-2)}l^2}{4b_0^2(d-1)} \left(\frac{d-2}{d-1}\frac{\pi b_1}{2b_0^2}+(2-d)c_0\right)m -\frac{V_{(d-2)}l^2 \beta^2 \gamma^2}{4b_0^2 (d-1)}\left(\frac{d-2}{d-1}\frac{\pi I_l}{2b_0^2}+c_2-(d-1)c_0\right)m \notag\\
 -\frac{V_{(d-2)}l^{d+2}}{(2b_0)^{d+2}}(v_{00} +\beta^2 \gamma^2 v_{01} -\beta^4 \gamma^4 v_{11})~.
\end{gather}
The holographic dual of the Fisher information metric can be defined as eq.$\eqref{deb}$ with the constant $C_d$
as follows
\begin{equation}
 C_d= \frac{a_1}{4G_{(d+1)}}\left(\frac{h_0 +h_1 \beta^2 \gamma^2 +h_2 \beta^4 \gamma^4 }{v_{00} +\beta^2 \gamma^2 v_{01} -\beta^4 \gamma^4 v_{11}}\right)~.
\end{equation}
It is obvious from the above expression that the constant $C_d$ matches with eq.$\eqref{cd0}$ in the $\beta \rightarrow 0$ limit. We would like to mention that the constant $C_d$ in this case depends on the boost parameter $\beta$ which in the case of the pure black brane  was independent of any physical parameter and depend only on the dimensionality of the spacetime.

\noindent We now look at the other holographic proposal \cite{MIyaji:2015mia} to compute the fidelity susceptibility.
For pure states the expression for fidelity \eqref{fidelity} reduces to 
\begin{equation}\label{purest}
\langle \Psi (\lambda) | \Psi (\lambda+ \delta \lambda) \rangle =1-G_{\lambda \lambda}  (\delta \lambda)^2+ \cdots.
\end{equation}
where for simplicity we assume that the states depend on a single parameter $\lambda$.
Therefore one can say that $G_{\lambda \lambda}$ measures the distance between two quantum states. 
$G_{\lambda \lambda}$ is called fidelity susceptibility . 
In $\cite{MIyaji:2015mia}$ it has been proposed that for a $d$- dimensional CFT deformed by a perturbation, the fidelity susceptibility can be computed holographically by the formula 
\begin{gather}{\label{fs}}
 G_{\lambda \lambda} = n_{d-1} \frac{Vol(\Sigma_{max})}{R^{d}}
\end{gather}
where $n_{d-1}$ is a $\mathcal{O}(1)$ constant and $R$ is the radius of curvature of $AdS$ spacetime.
$\Sigma_{max}$ is the maximum volume in the $AdS$ that ends at the $AdS$ boundary at a fixed time slice.
Though the above formula has been derived for the case of pure states, it has been also applied for mixed 
states \cite{MIyaji:2015mia}. Therefore we can apply this formula to calculate the 
fidelity susceptibility for boosted black brane.

\noindent Let us first calculate fidelity susceptibility for the $AdS$ black brane.
The metric for $AdS$ black brane in $(d+1)$- dimensions is given by
 \begin{eqnarray}
&&ds^2={1\over z^2}\left( -f dt^2+ dx_1^2+\cdots+dx_{d-1}^2+{dz^2 \over f}\right) 
\end{eqnarray}
with
\begin{equation}
  f=1-{z^d\over z_0^d} ~.
\end{equation}
Using the formula \eqref{fs}, the fidelity susceptibility reads
\begin{eqnarray}\label{fs1}
 G_{\lambda \lambda}&=& n_{d-1} L^{d-1} \int_{\delta}^{z_0} dz \frac{1}{z^d \sqrt{1-{z^d\over z_0^d}}} \notag\\
 &=&{n_{d-1} L^{d-1} \over z_0^{d-1}} \left[\frac{1}{d} B({1-d\over d},{1\over2}) +\frac{z_0^{d-1}}{(d-1)\delta^{d-1}} \right]~. \notag \\
\end{eqnarray}
We see that the above expression for the fidelity susceptibility does not agree with the Fisher information
metric obtained in eq.\eqref{fisher0}. We now proceed to calculate the fidelity susceptibility for the boosted black brane metric \eqref{bstmet}.
In this case the fidelity susceptibility takes the form 
\begin{eqnarray}
 G_{\lambda \lambda}&=& n_{d-1} L^{d-1} \int_{\delta}^{z_0} dz \frac{1}{z^d}\sqrt{\frac{K(z)}{f(z)}} \notag \\
 &=&n_{d-1} L^{d-1} \left[\frac{1}{z_0^{d-1}\sqrt{1-\beta^2}}\int_{0}^{1} dt \frac{1}{t^d}\sqrt{\frac{1-\beta^2(1- t^d) }{1-t^d}} +\frac{1}{(d-1)\delta^{d-1}}\right] 
\end{eqnarray}
where $t=z/z_0$. Now making a transformation $1-t^d = p$ yields
\begin{eqnarray}{\label{fs2}}
  G_{\lambda \lambda}&=& n_{d-1} L^{d-1} \left[\frac{1}{d \sqrt{1-\beta^2}z_0^{d-1}} \int_0^1 \frac{dp}{\sqrt{p}}\frac{\sqrt{1-\beta^2 p}}{(1-p)^{\frac{2d-1}{d}}} +\frac{1}{(d-1)\delta^{d-1}} \right]  \notag\\
  &=&n_{d-1} L^{d-1} \left[\frac{1}{d \sqrt{1-\beta^2}z_0^{d-1}}B({1\over 2},{1-d\over d}){}_2F_1(-\frac{1}{2},\frac{1}{2};\frac{2-d}{2d};\beta^2)+\frac{1}{(d-1)\delta^{d-1}}  \right]~.\notag\\
\end{eqnarray}
It is clear from eq. \eqref{fs2} that in the $\beta =0$ limit, the result matches with that of the $AdS$ black
brane given in eq. \eqref{fs1}. We see that the above expression for the fidelity susceptibility does not agree with the Fisher information
metric obtained in eq.\eqref{fisher1}. Although from quantum information literature we know that these two quantities are related (See appendix \eqref{reln}). A possible reason for this difference in the holographic results for the two metrics may be due to the difference in their definitions, which in the Fisher information metric case is an integration up to the turning point of the RT surface whereas in the fidelity susceptibility or the Bures metric case involves an integration up to the horizon radius of the black brane solution. The definition of fidelity susceptitbility given in \cite{MIyaji:2015mia} is an exact expression, while the Fisher information metric in \cite{Banerjee:2017qti} is obtained by computing the volume integral perturbatively upto second order. From the quantum information literature it is known that both the metrics are obtained by evaluating fidelity or relative entropy between a reference state ($\rho_0$) and a one parameter family of nearby states $(\rho= \rho_0+\lambda\delta\rho+\lambda^2\delta^2\rho)$.
It should be noted that the expressions of the two metrics from the quantum information perspective gets contribution from the first order perturbation in the density matrix
eq.\eqref{FI}, Appendix \eqref{reln} whereas the expressions from the bulk require perturbations upto second order in the bulk metric and perturbations of the extremal surface. However it is not clear how this perturbation of density matrices on the boundary are exactly related to the perturbations of the extremal surface and the metric in the bulk. This discrepancy could be a reason for the two results not being in agreement with the results from the quantum information viewpoint. 

Further, from the bulk perspective we see that for the Fisher information metric case an integration up to the turning point of the RT surface is involved which is obtained perturbatively. This act of evaluating the volume integral perturbatively contains the information of the asymptotic region only \cite{Bhattacharya:2012mi,Graham:1999pm}. However, the fidelity susceptibility or the Bures metric case involves an integration up to the horizon radius of the black brane solution and the integral is evaluated exactly, no ordering is maintained. Thus it contains the information of the full spacetime geometry. This could be another reason for the disagreement.

\section{Conclusion}{\label{5}}
In this paper, we have computed the holographic subregion complexity for a boosted black brane for two cases where the boost direction is in the parallel and perpendicular direction to the strip length comprising the subsystem. The computation have been carried out upto both first and second orders in the boost parameter. An asymmetry has been found in the holographic subregion complexity upto first order. 
This asymmetry in the holographic subregion complexities owes its origin to the asymmetry in the holographic entanglement entropies in the perpendicular and the parallel directions which in turn can be related to the unequal entanglement pressure.
The results upto second order in the boost parameter have been used to compute the Fisher information metric upto an undetermined constant. This constant has been fixed by equating this result with the holographic computation of Fisher information metric from the relative entropy. This has been done for both the pure black brane as well as the boosted black brane. The fidelity susceptibility has also been obtained by using the proposal in \cite{MIyaji:2015mia}. It is observed that the expressions for the Fisher information metric and the fidelity susceptibility are not related to each other \cite{Dominik}. This is one of the main results in this paper and shows a remarkable dissimilarity with the results present in the quantum information literature where it is known that the two distances may be related in general. It would be interesting to investigate in what bulk limit or approximation does the fidelity susceptibility gets related to the Fisher information metric. 

\appendix
\section{List of Beta function Identities}\label{betas}
In this appendix we give some useful Beta function integrals which we have used in the paper. 
\begin{eqnarray}
&& b_0=\int_0^{1} dt\; t^{d-1}{1 \over   \sqrt{R}}
={1\over 2(d-1)}B({d\over2d-2},{1\over2})=\frac{\sqrt{\pi } \Gamma \left(\frac{d}{2 d-2}\right)}{\Gamma \left(\frac{1}{2 (d-1)}\right)}\notag\\
&& b_1=\int_0^{1} dt\; t^{2d-1}{1 \over   \sqrt{R}}
={1\over 2(d-1)}B({d\over d-1},{1\over2})=\frac{\sqrt{\pi } \Gamma \left(\frac{d}{d-1}\right)}{(d+1) \Gamma \left(\frac{1}{2}+\frac{1}{d-1}\right)}\notag\\
&& b_2=\int_0^{1} dt \;t^{3d-1}{1 \over   \sqrt{R}}
={1\over 2(d-1)}B({3d\over 2d-2},{1\over2})\notag\\
&& I_l=\int_0^{1} dt\; t^{d-1}(1-t^d){1 \over {R}^{3\over2}}=
{d+1\over d-1} b_1
-{1\over d-1} b_0 \notag\\
&&c_0=\int^1_0 dt {t^d \over\sqrt{R}}=\frac{1}{2(d-1)}B({d+1\over 2(d-1)},{1\over2})=\frac{\pi}{2(d^2-1)b_1} \notag\\
&&c_1=\int^1_0 dt {t^{2d} \over\sqrt{R}}=\frac{1}{2(d-1)}B({2d+1\over 2(d-1)},{1\over2})=\frac{\pi}{2(2d+1)(d-1)b_2} \notag\\
&&c_2=\int^{1}_{0} dt{{(1-t^d)\over{R}^{3\over 2}}}=\frac{2}{d-1}c_0 +\frac{d-2}{2(d-1)^2}B({1\over 2(d-1)},{1\over2})=\frac{\pi}{(d+1)(d-1)^2 b_1}
+ \frac{\pi (d-2)}{2(d-1)^2 b_0}\notag\\
\notag\\
&& c_3 = \int^1_0 dt \frac{t^d(1-t^d)}{R^{3 \over 2}}=\frac{2}{d-1}c_0 + \frac{d+2}{d-1}c_1 \notag\\
&&\int^{1}_{0}{dt\over\sqrt{1-t^{2(d-1)}}}={\pi\over 2(d-1)b_0}\notag\\
&& J_l=\int_0^{1} dt\; t^{d-1} \left( 
{\beta^2\gamma^2\over 4} t^d
+\beta^4\gamma^4\left(\frac{3(1-t^d)}{8(1-t^{2(d-1)})}-\frac{1}{2} \right) \right)
{(1-t^d) \over   {R}^{3\over2}}= \beta^2\gamma^2 J_1 + \beta^4\gamma^4 J_2\notag\\
&& K_l=\int_0^{1} dt  \left( 
{\beta^2\gamma^2\over 4} t^d
+\beta^4\gamma^4 \left(\frac{3(1-t^d)}{8(1-t^{2(d-1)})}-\frac{1}{2}\right)\right)
{(1-t^d) \over   {R}^{3\over2}} = \beta^2\gamma^2 K_1 - \beta^4\gamma^4 K_2 
\end{eqnarray}
with 
\begin{eqnarray}
 &&J_1 = \frac{1}{4(d-1)}\left((2d+1)b_2 -(d+1)b_1\right) \notag \\
 && J_2= \frac{1}{8(d-1)^2}\left((3-2d)b_0 -2(d+1)(3-d)b_1+3(2d+1)b_2 \right) -\frac{I_l}{2} \notag \\
 && K_1 =\frac{1}{2(d-1)}c_0 +\frac{d+2}{4(d-1)}c_1 \notag \\
 && K_2 =-\frac{d-4}{4(d-1)^2}c_0+\frac{(d+2)(d-4)}{8(d-1)^2}c_1+\frac{d}{8(d-1)}c_2
\end{eqnarray}

where $B(m,n)={\Gamma(m)\Gamma(n)\over\Gamma(m+n)}$ are the Beta-functions and we have used the identity
$B(x,\frac{1}{2}) B(x+\frac{1}{2},\frac{1}{2})=\frac{\pi}{x}$.
Further integrals are
\begin{eqnarray}
&& a_0=\int_0^{1} dt\; t^{-d+1}{1 \over   \sqrt{R}}
={1\over 2(d-1)}B({1-d/2\over d-1},{1\over2})\notag\\
&& a_1=\int_0^{1} dt \;t^{-d+1}{t^d \over   \sqrt{R}}
={1\over 2(d-1)}B({1\over d-1},{1\over2})\notag\\
&& a_2=\int_0^{1} dt\; t^{-d+1}{t^{2d} \over   \sqrt{R}}
={1\over 2(d-1)}B({1+d/2\over d-1},{1\over2})\notag\\
&& I_a=\int_0^{1} dt\; t^{d-1}(1-t^{2d}){1 \over   {R}^{3/2}}
={2d+1\over d-1} b_2
-{1\over d-1} b_0 ~.
\end{eqnarray}
Some identities we  have used are 
\begin{eqnarray}
b_0=(2-d) a_0, ~~~~
b_1={2\over d+1} a_1, ~~~~
b_2={2+d\over 2d+1} a_2 \ .
\end{eqnarray}


\section{Relation between quantum Fisher information and the Bures metric}\label{reln}
Here we will review the relation between the Bures metric and the quantum Fisher information metric. Our discussion here will follow that of \cite{hubner1992explicit}-\cite{Dominik}. The Bures distance is a measure of distinguishability between two quantum states $\rho_1$ and $\rho_2$ and is defined as
\begin{equation*}
 D_B(\rho_1,\rho_2)=2\sqrt{1-F(\rho_1,\rho_2)}
\end{equation*}
where $F(\rho_1,\rho_2)$ is the fidelity between the two states and is defined as
\begin{equation}\label{fidel}
 F(\rho_1,\rho_2)=Tr\sqrt{\sqrt{\rho_1}~\rho_2~\sqrt{\rho_1}}~.
\end{equation}
Now the Bures metric \cite{bures1969extension} is defined as
\begin{equation}\label{bumetric}
 ds^2_B=g_{ij}d\lambda_id\lambda_j=D_B(\rho,\rho+\lambda\rho)^2~.
\end{equation}
If $\rho$ and $\sigma$ are both pure states then the Bures metric 
reduces to eq.\eqref{purest}.

Another important metric on the state space is the quantum Fisher information metric \cite{paris2009quantum}.  It is a symmetric positive or positive semidefinite metric given by
\begin{equation}\label{fish}
 H_{ij}={1\over 2}Tr\left[\left(L_iL_j+L_jL_i\right)\rho\right]
\end{equation}
where $L_i$ is the symmetric logarithmic derivative and is defined as the operator solution to the equation
\begin{gather*}
 {1\over 2}\left(L_i\rho+\rho L_i\right)=\partial_{i}\rho~.
\end{gather*}
Using the spectral decomposition of the density matrix as $\rho=\sum_k p_k\ket{k}\bra{k}$ one can rewrite eq.\eqref{fish} as
\begin{equation}\label{heris}
 H_{ij}=2\sum_{p_k+p_l>0}\frac{Re(\bra{k}\partial_{i}\rho\ket{l}\bra{l}\partial_{j}\rho\ket{k})}{p_k+p_l}
\end{equation}
where $Re$ denotes the real part. The statistical equivalent distance of this metric is obtained by maximizing the mixed state generalization of the Fisher information over all possible quantum measurements \cite{PhysRevLett.72.3439}. Now in order to find a relation between the two metrics we need to calculate the Bures distance between two infinitesimally close states $\rho_1=\rho_0$ and $\rho_2=\rho_0+\lambda\delta\rho$. In order to calculate this we need to first obtain an expression for $F(\rho_0,\rho_0+\lambda\delta\rho)$. Let us consider the expression
\begin{equation}\label{kyzs}
 \sqrt{\sqrt{\rho_0}~(\rho_0+\lambda\delta\rho)~\sqrt{\rho_0}}=\rho_0+X+Y
\end{equation}
where $X$ and $Y$ are the first and second order terms in $\lambda$. Now squaring the above equation and equating terms at each order in $\lambda$ gives
\begin{equation*}
 \sqrt{\rho_0}~\delta\rho~\sqrt{\rho_0}=X\rho_0+\rho_0X~,~ -X^2=Y\rho_0+\rho_0Y~.
\end{equation*}
One can write the above expression in a diagonal basis of $\rho$ with eigenvalues $p_k>0$ as
\begin{equation*}
 \bra{k}X\ket{l}={p^{1\over 2}_k p^{1\over 2}_l\over (p_k+p_l)}\bra{k}\delta\rho\ket{l},~~~~~~ \bra{k}Y\ket{l}={- \bra{k}X^2\ket{l}\over (p_k+p_l)}~.
\end{equation*}
Since $Tr\rho=1$ thus $Tr=\delta\rho=0$. This implies that $TrX=0$ and hence the only contribution comes from the second order term. The trace of the second order term is 
\begin{equation}\label{kata}
 TrY=-{1\over 4}\sum_{p_k+p_l>0}{\mid\bra{k}\delta\rho\ket{l}\mid^2\over (p_k+p_l)}~.
\end{equation}
Now from eq.(s)\eqref{fidel}, \eqref{kyzs} we can write the fidelity as 
\begin{equation}\label{kol}
F(\rho_0,\rho_0+\lambda\delta\rho)=Tr\rho_0+TrX+TrY=1+TrY~.
\end{equation}
Thus the Bures distance from eq.\eqref{bumetric} is given by
\begin{equation*}
 ds^2_B=g_{ij}d\lambda_id\lambda_j=D_B(\rho_0,\rho_0+\lambda\delta\rho)^2=2\left(1-F(\rho_0,\rho_0+\lambda\delta\rho)\right)
\end{equation*}
and thus using eq.(s)\eqref{kol}, \eqref{kata} and \eqref{heris}, we get 
\begin{equation}\label{rel}
 g_{ij}={1\over 2}\sum_{p_k+p_l>0}{Re(\bra{k}\partial_{i}\rho\ket{l}\bra{l}\partial_{j}\rho\ket{k})\over (p_k+p_l)}={1\over 4}H_{ij}~.
\end{equation}
The above expression gives the relation between Bures and quantum Fisher information metric. It is important to note that the expression \eqref{rel} was obtained by considering only first order terms in the density matrix perturbation. On accounting for second order perturbation, eq.\eqref{rel} will receive contributions from $\rho$ and its derivatives \cite{Dominik}.
\section*{Acknowledgment}
SG acknowledges the support of IUCAA, Pune for the Visiting Associateship programme. RM acknowledges Harvendra Singh for helpful discussions. 
The authors would like to thank the referee for very useful comments which improved the discussion in the paper.

\end{document}